\documentclass[12pt]{article}
\usepackage{amsmath,amsfonts}
\usepackage{graphicx}
\def\bra{\langle}
\def\ket{\rangle}
\def\d{\partial}

\newcommand{\R}{\mathbb{R}}

\def\lr{{L^2(\R)}}

\def\dk#1#2{\frac{ d^{#2}{#1} }{ (2\pi)^{#2} }} % invariant measure in FT

\def\eg{{\sl e.g. }}
\def\etc{{\sl etc.}}
\def\ie{{\sl i.e. }}

\def\vx{\mathbf{x}}
\def\ve{\mathbf{e}}
\def\vu{\mathbf{u}}
\def\vb{\mathbf{b}}
\def\vk{\mathbf{k}}
\def\vp{\mathbf{p}}
\def\vq{\mathbf{q}}

\def\veta{\mathbf{\eta}}
\begin{document}
\title{Multiscale theory of turbulence in wavelet representation}
\author{M.V.Altaisky \\ Joint Institute for Nuclear Research, Dubna, 141980, Russia;\\ 
and Space Research Institute, Profsoyuznaya 84/32, \\ Moscow, 117997,
Russia; e-mail: altaisky@mx.iki.rssi.ru}
\date{June 7, 2004}
\maketitle
\begin{abstract}
We present a multiscale description of hydrodynamic turbulence
in incompressible fluid based on a continuous wavelet transform (CWT) and 
a stochastic hydrodynamics formalism. 
Defining the stirring random force by the correlation function of its 
wavelet components, we achieve the cancellation of loop divergences in 
the stochastic perturbation expansion. 
An extra
contribution to the energy transfer from large to smaller scales is
considered. It is shown that the Kolmogorov hypotheses are naturally
reformulated in multiscale formalism. The multiscale perturbation
theory and statistical closures based on the wavelet decomposition 
are constructed.
\end{abstract}
\leftline{PACS: 47.27.Gs, 11.10.Gh}
\section{Introduction}
The statistical description of a fully developed hydrodynamic turbulence
is based on the Kolmogorov  hypotheses \cite{K41a} on the
self-similarity of the velocity fluctuations of different scales.
However, the Kolmogorov-Obukhov  analysis \cite{K1962,Ob1962} 
does not provide 
a rigorous mathematical definition of the ``fluctuation of scale
$l$''. In the literature this is tacitly understood as 
Fourier components with wavenumbers approximately equal to the inverse 
scale $k\!\approx\!\frac{2\pi}{l}$ and the
analysis is performed in wavenumber space. This definition meets
global characteristics of the fully developed isotropic turbulence,
but, being based on the Fourier transform, is essentially nonlocal
and therefore hardly applicable to such important properties of the 
fully developed turbulence as the coherent structure formation. To
catch the local properties of the turbulent velocity field, the
decomposition into localized wave packets
and wavelets have been performed by many authors
\cite{Zimin1981,Nakano1988,Meneveau1991,VerFri1991,Farge1992,Ast1996}.

Most of the wavelet applications to turbulence are
restricted either to an analysis of the measured  turbulent fields 
with the ``wavelet microscope'', capable of simultaneous analysis
of the same velocity field at different resolution
\cite{VerFri1991,Farge1992,Ast1996}, or to a numerical solution
of the Navier-Stokes equations (NSE) in the wavelet basis
\cite{Meneveau1991,Reslea1995,GL1996,FS1998}.  
The application of the continuous wavelet transform with the 
derivatives of the Gaussian 
taken as basic wavelets was already used in the analytical study of 
the NSE \cite{Lewalle1993pf,Lewalle1993as}, in particular,  
in research on energy dissipation. However, at least to the authors 
knowledge {\em the wavelet decomposition has not been yet applied 
to the stochastic iterative solutions of the Navier-Stokes equation 
or in the framework of the field theory approach to statistical hydrodynamics}
\cite{FNS1977,YO1986,AAV1999}. 

Our interest in extending the stochastic hydrodynamics approach by 
wavelet-defined random processes is stimulated by recent developments 
in application of the field theory and renormalization group methods to 
the fully developed isotropic turbulence, see e.g. \cite{GKV2002} for a short 
summary. The ``field theoretic methods'' are taken to mean the stochastic 
diagram technique, the functional integral representation of the 
characteristic functional and the renormalization group methods. The latter,
being inherited from the theory of critical behavior in equilibrium 
phase transitions clearly demonstrates the need for a scale-dependent 
random measure in the field theoretic approach to turbulence. Such measures, 
having been already in use for phenomenological studies of multifractal 
behavior and itermittency \cite{ABM1991,VerFri1991,Farge1992}, have not 
been yet used in analytical approach of stochastic hydrodynamics.  

The aim of this paper is to extend the  wavelet 
representation of the stochastic Navier-Stokes equations in 
such a way that the probability distribution of the stirring 
force, used to compensate the energy dissipation, is defined for the 
wavelet coefficients of random force  
(\ie for the scale components of forcing).  In our  
description the velocity field wavelet
coefficients $u_l(x)$ attain the Kolmogorov meaning of local 
velocity fluctuations of scale $l$ at a given point $x$ 
(A particular case 
of the difference of two Gaussians or delta-functions was  
considered in the literature \cite{VerFri1991,Arneodo:book}). 
Defining the random force in the space of wavelet coefficients we 
get an extra analytical flexibility: there are   
random processes in the space of wavelet coefficients  with different 
correlation functions whose images 
under inverse WT coincide in the space of common 
random functions. Tuning the random force 
correlation function in the space of wavelet coefficients, 
in stochastic hydrodynamics formalism, we get rid of loop 
divergences; for a special type 
of narrow-band forcing  the contributions 
to the response and the correlation functions are explicitly calculated.    

It is shown that the Kolmogorov hypotheses, the statistical closures of moment 
equations, the stochastic hydrodynamics approach and the Wyld diagram 
technique  are
naturally reformulated in multiscale (wavelet) formalism. Besides, the 
consideration of random processes  depending on scale
explicitly $\bra u_l(x) u_{l'}(x')\ket = C(x,x',l,l')$ gives a
possibility of the perturbation expansion converging without
introducing an ultra-violet (UV) cutoff wavenumber, and an extra contribution to the
Kolmogorov energy dissipation term  $(u_l^3/l)$.

The rest of this paper is organized as follows. In {\em
Section 2} we review the methods of solution of the NSE by the Fourier
decomposition and by the wavelet decomposition. In {\em Section 3} the
stochastic hydrodynamics approach is reformulated for the random
processes explicitly depending on scale. A regularization of the
perturbation expansion for a random force acting at a single
scale is presented. Energy dissipation rate and energy flux in the
multiscale formalism are considered in {\em Section 4}. {\em
Section 5} presents a generalized form of the Kolmogorov
hypotheses formulated in a multiscale framework. In {\em Conclusion}
we discuss some perspectives of the method.

\section{Hydrodynamics of incompressible fluid in wavelet representation}
In analytical studies of the hydrodynamic turbulence the basic
role is played by the Navier-Stokes equations:
\begin{equation}
\frac{\d \vu}{\d t}+ (\vu\cdot\nabla) \vu = \nu \Delta \vu
                    -\nabla\left(\frac{p}{\rho}\right) .
% + f(\vx,t)
\label{nse}
\end{equation}
%where $\nu$ is kinematic viscosity and $f(\vx,t)$ is the  
%regular force applied to the fluid.
%%%%%%%%%%%%%%%%%%%%%%%%%%%%%%%%%%%%%%%% 26.05.04 %%%%%%%%%%%%
Our consideration of hydrodynamic turbulence, based on the Navier-Stokes 
equations, assumes a fully developed homogeneous isotropic turbulence 
far from any boundaries. Although the adequacy of the stochastic NSE 
to fully developed turbulence still remains an open problem, a significant 
progress has been achieved in studying simple models, 
proving the Kolmogorov spectrum from basic principles \cite{MTY1976} and 
calculating the anomalous scaling corrections \cite{CFKL1995,GK1995,AHHJ2003}. 
Turbulence in an incompressible fluid considered in these settings is more 
frequently studied in a wavenumber space rather than in a real space. The 
advantage of wavenumber space is a very simple form of the Laplacian -- this  
enables to eliminate the pressure term from the NSE. 
The price paid for this simplification is the non-locality of the Fourier 
transform that completely hides all information related to the real space 
distribution of the velocity field. The discrete Fourier transform used 
in numerical simulations also imposes periodicity on the system. The 
usefulness of the wavenumber representation stems from two basic reasons: 
the existing of fast FFT algorithms and pseudospectral methods 
\cite{OrsPat1972,BSM1999} and the direct experimental interpretation of the 
power spectra density of velocity fluctuations \cite{Frish1995}. 
%%%%%%%%%%%%%%%%%%%%%%%%%%%%%%%%%%%%%%%%%%%%%%%%%%%%%%%%%%%%%% 
For the incompressible  fluid  the pressure term can be eliminated
from the NSE \eqref{nse} by the substitution
\begin{equation}
p=-\frac{\rho}{\Delta}\left(\frac{\d u_i}{\d x_j}\frac{\d u_j}{\d
x_i} \right). \label{pexc}
\end{equation}
In the wavenumber space representation 
\begin{equation}
u_i(\vx,t) = \int e^{\imath \vk \vx}  \hat u_i(\vk,t)
\dk{\vk}{d},\quad u_i(\vx,t) = \int e^{\imath (\vk \vx -\omega t)}
\hat u_i(\vk ,\omega) \dk{k}{d+1},\quad \label{ft}
\end{equation}
both the inverse Laplacian operator $\Delta^{-1}$ in \eqref{pexc} 
and the incompressibility condition $\nabla\!\cdot\!\vu=0$
are simplified and the NSE system becomes 
a system of integro-differential equations 
\begin{eqnarray}
\left(\d_t + \nu\vk^2 \right)\hat u_i(\vk,t) &=&
\int \dk{\vq}{d} M_{ijk}(\vk) \hat u_j(\vq,t)\hat u_k(\vk-\vq,t),\label{feq1}\\
\nonumber \hbox{where}\quad M_{ijk}(\vk) &=& - \frac{\imath}{2}\left[
                   k_j P_{ik}(\vk)+ k_k P_{ij}(\vk)
                                   \right] , \quad
P_{ij}(\vk)  = \delta_{ij} - \frac{k_i k_j}{\vk^2}.
\end{eqnarray}

The system of equations \eqref{feq1} allows for perturbative
calculations, statistical closures of moment equations \etc., but is not 
local in the 
coordinate space and therefore is of little use when 
studying the effects locally produced by
fluctuations. The system of equations \eqref{feq1} is complete
and, being correctly solved numerically, gives reliable results.
This however requires taking into account a tremendous number of
Fourier modes, that is hardly bearable even for modern
supercomputers; alternatively certain block-averaging 
procedures in wavenumber space can be applied, \eg the spectral reduction \cite{BSM1999}.

The idea of studying turbulence using the space-scale or wavenumber-scale 
representation is not a new one. 
The band-pass filtering of the turbulent signals 
was used to study intermittency long ago \cite{KenCor1961}. Later the 
prototype of wavelet cascade model was given in \cite{Zimin1981}. 
The wavelet transform (WT) is known to be an excellent local tool, widely used
in a turbulence data analysis \cite{Farge1992} and numerical
simulations \cite{Zimin1981,Meneveau1991,FS1998}. WT in real 
space, as a particular type of filtering \cite{Germano1992}, reveals the 
formation of the coherent structures \cite{Farge1992}  and  
is useful to study the local energy dissipation effects related 
to filament formation \cite{GL1996}. WT in $(a,k)$ 
space enables to enhance the idea of band-pass filtering by 
separating the contributions of different scales.  

In this paper we use continuous wavelet transform (CWT)  
to derive the equations for the fluctuations of different scales. This 
analytical consideration goes along with the filtering approach 
\cite{Germano1992,Lewalle1994}, when considering moment closures 
for a fully developed isotropic turbulence.  
{\em The crux of our approach is the application of continuous wavelet 
transform 
in the stochastic hydrodynamics framework: the WT is applied to both the 
velocity field and the random force stirring the turbulence}. Working 
in the space of wavelet coefficients, rather than in a real space of 
velocity fields, we can define the random stirring force, which 
is essentially used in RG calculations, see 
\eg \cite{McComb,AAV1999} -- in  a way that 
allows for getting rid of loop divergences in the stochastic perturbation 
expansion of the velocity field statistical momenta \cite{AltDAN03e}.

For simplicity we restrict 
ourselves with the homogeneous isotropic turbulence and the isotropic wavelets
$\psi(\vx)=\psi(|\vx|)$. In this case, the wavelet transform of the
velocity field $\vu(x,t)$, taken with respect to the basic wavelet
$\psi(\vx)$, and the corresponding reconstruction formula are
\begin{eqnarray}
\vu_a(\vb,t) &=& \int \frac{1}{|a|^d}
               \bar\psi\left(\frac{\vx-\vb}{a} \right)\vu(\vx,t) d^d \vx,
               \label{dcwt} \\
\vu(\vx,t)   &=& \frac{1}{C_\psi} \int 
\psi\left(\frac{\vx-\vb}{a} \right) \vu_a(\vb,t)
\frac{da d^d{\vb}}{|a|^{d+1}}. \label{icwt}
\end{eqnarray}
We perform the wavelet transform only in the spatial argument of the
velocity field because we need the {\em spatial} resolution. Using the $L^1$ 
norm instead of $L^2$, we provide wavelet coefficients
$\vu_a(\vb,t)$ with the same $(LT^{-1})$ dimension as the 
velocity field $\vu(x)$ itself. The wavelet coefficients $\vu_a(\vb,t)$
are referred to hereafter, as the components of the velocity
field corresponding to  scale $a$; $\psi(x)$ is referred to as an
analyzing function used to measure the scale components.

For practical calculations it is often convenient to express
wavelet transform (\ref{dcwt},\ref{icwt}) in $(a,\vk)$
representation taking the Fourier transform in the spatial
argument. In Fourier form the direct and inverse WT are:
\begin{equation}
\hat\vu_a(k) = \overline{\hat\psi(a\vk)}\hat\vu(k), \quad
\hat\vu(k)   = \frac{1}{C_\psi} \int \frac{da}{|a|}
\hat\psi(a\vk)\hat\vu_a(k), \label{wtf}
\end{equation}
     where $\hat\vu(k)\equiv \hat\vu(\vk,\omega)$ is the Fourier
		   transform of the velocity field
$$u(\vx,t) = \int e^{\imath(\vk\vx-\omega t)}u(k)\frac{d^d\vk
d\omega}{(2\pi)^{d+1}}.$$ 
(The Minkovski-like notation $x\equiv(\vx,t),k\equiv(\vk,\omega)$ is used.)
Therefore, the wavelet
transform \eqref{dcwt} can be considered as a frequency filter  
that conveys the harmonics with typical wavenumbers of order
$\frac{1}{a}$ and is localized close to point $\vb$.

The only restriction imposed on the basic wavelet $\psi$ to make the 
wavelet transform invertible -- the admissibility condition -- is
the finiteness of the normalization constant $C_\psi$:
\begin{equation}
C_\psi = \int_{-\infty}^{\infty} \frac{|\hat\psi(a\vk)|^2}{|a|}da  < \infty.
\label{adc1}
\end{equation}
For a real-valued basic wavelet $\psi(\vx)$ we can restrict the integration 
to the positive frequencies only 
$$
C_\psi = 2\int_0^\infty \frac{|\hat\psi(a\vk)|^2}{a}da.$$
If the basic wavelet is also isotropic $\psi(\vx)=\psi(|\vx|)$, we get 
\begin{equation}
C_\psi = \int_{\R^d} \frac{|\hat\psi(\vk)|^2}{S_d|\vk|^d}d^d\vk,
\label{adc2}
\end{equation}  
where $S_d$ is the area of the unit sphere in $\R^d$.

In this paper we will assume $a\!\in\!\R_+$ integration and the isotropic 
real wavelets. Thus, the decomposition of the velocity field 
with respect to the basic wavelet $\psi$ takes the form 
\begin{eqnarray}
u(\vx,t) &=& \frac{2}{C_\psi} \int_0^\infty \frac{da}{a^{d+1}}
\int_{\R^d}d^d\vb \psi \left(\frac{\vx-\vb}{a} \right)\vu_a(\vb,t) \label{wfp}\\
u(\vx,t) &=& \frac{2}{C_\psi} \int_0^\infty \frac{da}{a} 
\int_{\R^d}\frac{d^d\vk d\omega}{(2\pi)^{d+1}}
e^{\imath (\vk\vx-\omega t)} \hat\psi(a\vk) \hat u_a(\vk,\omega) . \label{wfs}
\end{eqnarray}
We drop the integration limits $\int_0^\infty \frac{da}{a}$ hereafter. 
%%%%%%%%%%%%%%%%%%%%%%%% 31.05 %%%%%%%%%%%%%%%%%%%%%%%%%%%%%%%%%   

Substituting the wavelet transform \eqref{wfs} into the system of
the component equations \eqref{feq1}, we yield the system of
equations for the scale components $\hat\vu_{ai}(k)$:
\begin{eqnarray}
\nonumber (-\imath\omega+\nu\vk^2)\hat
u_{ai}(k)&=&\left(\frac{2}{C_\psi}\right)^2\int
M_{ijk}^{aa_1a_2}(\vk,\vq,\vk\!-\!\vq) \hat u_{a_1j}(q)\hat
u_{a_2k}(k\!-\!q)
\frac{da_1}{a_1}\frac{da_2}{a_2}\frac{d^{d+1}q}{(2\pi)^{d+1}} \\
M_{ijk}^{aa_1a_2}(\vk,\vq,\vk\!-\!\vq) &=&
\overline{\hat\psi(a\vk)}M_{ijk} (\vk)\hat\psi(a_1\vq)
\hat\psi(a_2(\vk\!-\!\vq)). \label{wak}
\end{eqnarray}
Let us derive statistical closures for the scale components. For this
purpose we take Eq.\eqref{wak} and its complex conjugate 
\begin{equation}
\begin{array}{lcl}
(\d_t+\nu\vk^2) \hat u_{ai}(\vk,t) &=& \left(\frac{2}{C_\psi}\right)^2
\int
M_{ijk}^{aa_1a_2}(\vk,\vq,\vk\!-\!\vq) \hat u_{a_1j}(\vq,t) \hat
u_{a_2k}(\vk\!-\!\vq,t)
\frac{da_1}{a_1}\frac{da_2}{a_2}\frac{d^{d}\vq}{(2\pi)^{d}} \\
(\d_{t'}+\nu\vk^2) \overline{\hat u_{ai}(\vk,t')} &=& 
\left(\frac{2}{C_\psi}\right)^2\int
\overline{M_{ijk}^{aa_1a_2}(\vk,\vq,\vk\!-\!\vq) \hat
u_{a_1j}(\vq,t') \hat u_{a_2k}(\vk\!-\!\vq,t')}
\frac{da_1}{a_1}\frac{da_2}{a_2}\frac{d^{d}\vq}{(2\pi)^{d}},
\end{array}
\label{wat}
\end{equation}
multiply the first equation by 
$\overline{\hat u_{ai}(\vk,t')}$, sum up over the vector index 
$i$ and take the statistical averaging $\langle\ \rangle$. Doing so, we get
\begin{eqnarray*}
(\d_t+\nu\vk^2)\sum_i \bra 
\overline{\hat u_{ai}(\vk,t')} u_{ai}(\vk,t) \ket &=& \left(\frac{2}{C_\psi}\right)^2
\int \frac{da_1}{a_1}\frac{da_2}{a_2}\frac{d^{d}\vq}{(2\pi)^{d}}
M_{ijk}^{aa_1a_2}(\vk,\vq,\vk\!-\!\vq) \\
 & & \bra \overline{\hat u_{ai}(\vk,t')} \hat u_{a_1j}(\vq,t)
\hat u_{a_2k}(\vk\!-\!\vq,t)\ket .
\end{eqnarray*}
Applying the same procedure to the second of the equations
\eqref{wat} and summing up the results at coinciding time arguments 
$t\!=\!t'$, we get the moment 
equation 
\begin{eqnarray} \displaystyle
\nonumber (\d_t+2\nu\vk^2)\sum_i \bra \overline{\hat
u_{ai}(\vk,t)}u_{ai}(\vk,t)  \ket &=& \left(\frac{2}{C_\psi}\right)^2 \int
\frac{da_1}{a_1}\frac{da_2}{a_2}\frac{d^{d}\vq}{(2\pi)^{d}}
M_{ijk}^{aa_1a_2}(\vk,\vq,\vk\!-\!\vq) \\
 & & \bra \overline{\hat u_{ai}(\vk,t)} \hat u_{a_1j}(\vq,t) \hat u_{a_2k}(\vk\!-\!\vq,t)\ket
 + h.c.,
\label{clos23}
\end{eqnarray}
which is different from its plane wave counterpart only by extra 
scale indexes and extra
integrations in scale logarithms $\int\frac{da}{a}$(octaves). To express the third
order moments in \eqref{clos23} via the second
moments, we must substitute
\begin{equation}
\begin{array}{lcl}
\displaystyle \hat u_{ai}(\vk,t) &=& \left(\frac{2}{C_\psi}\right)^2
\int_{-\infty}^t ds G_{il}^{aa_0}(\vk,t-s) M_{ljk}^{a_0 a_1
a_2}(\vk,\vq,\vk\!-\!\vq) \\
& &   \hat u_{a_1j}(\vq,s) \hat u_{a_2k}(\vk\!-\!\vq,s)
\frac{da_0}{a_0}\frac{da_1}{a_1}\frac{da_2}{a_2}\frac{d^{d}\vq}{(2\pi)^{d}},
\label{gfd16}
\end{array}
\end{equation}
where $G_{il}^{aa_0}(\vk,t\!-\!s)$ is the response function. 
The difference from the standard plane-wave
approach \cite{Nakano1988} is that additionally to the summation
over vector indices we have to sum up over octaves to 
integrate over $\int\frac{da}{a}$ in each scale variable.
So the statistical closures can be reproduced for the scale components.

In the zero-th order approximation (with no interaction term:
$M(\cdot)\to0$) the bare response function is given by 
\begin{equation}
G_{il}^{[{\rm bare}]aa_0}(\vk,t\!-\!s) = \delta_{il} \int_{-\infty}^\infty
\frac{\delta(a-a_0)a_0}{-\imath\omega+\nu\vk^2}e^{-\imath\omega(t-s)}\frac{d\omega}{2\pi}
= \delta_{il} \delta(a-a_0)a_0 e^{-\nu\vk^2|t-s|}. \label{brf}
\end{equation}
The full response function, in view of the component equations
\eqref{wat}, satisfies the integro-differential equation
$$
(\d_t+\nu\vk^2)G_{il}^{aa_0}(\vk,t-s)  = 2\left(\frac{2}{C_\psi}\right)^2\int
M_{ijk}^{aa_1a_2}(\vk,\vq,\vk\!-\!\vq) \hat u_{a_1j}(\vq,t)
G_{kl}^{a_2a_0}(\vk\!-\!\vq,t-s)
\frac{da_1}{a_1}\frac{da_2}{a_2}\frac{d^{d}\vq}{(2\pi)^{d}}.
$$
The substitution of \eqref{gfd16} into \eqref{clos23} gives a 
relation between the second and the forth order moments of the scale
components:
\begin{eqnarray} \displaystyle
\nonumber (\d_t+2\nu\vk^2)\sum_i \overline{\bra \hat
u_{ai}(\vk,t)}\hat u_{ai}(\vk,t)  \ket = 2\left(\frac{2}{C_\psi}\right)^4
\int
\frac{da_1}{a_1}\frac{da_2}{a_2}\frac{d^{d}\vk_1}{(2\pi)^{d}}
\frac{da_3}{a_3}\frac{da_4}{a_4}\frac{d^{d}\vk_2}{(2\pi)^{d}} \\
\nonumber\frac{da_0}{a_0}
M_{ijk}^{aa_1a_2}(\vk,\vk_1,\vk\!-\!\vk_1) \bra \overline{\hat
u_{ai}(\vk,t)} \hat u_{a_1j}(\vk_1,t)
\int_{-\infty}^t ds G_{kl}^{a_2a_0}(\vk\!-\!\vk_1,t,s) \\
M_{lrf}^{a_0a_3a_4}(\vk\!-\!\vk_1,\vk_2,\vk-\vk_1-\vk_2) \hat
u_{a_3r}(\vk_2,s) \hat u_{a_4f}(\vk\!-\!\vk_1\!-\!\vk_2,s) \ket
 + h.c..
\label{clos24}
\end{eqnarray}
The forth order moments $\bra uuuu \ket$ can be further decomposed
into the sum of all pairs $\bra uu \ket \bra uu \ket$ using
a stochastic perturbation expansion.

\section{Stochastic hydrodynamics with multiscale forcing}
The stochastic hydrodynamics approach consists in introducing random force 
in the Navier-Stokes equations and calculating the 
velocity field momenta $\bra \vu(x_1)\ldots\vu(x_n) \ket$ using the 
stochastic perturbation theory, pioneered by Wyld \cite{Wyld1961}, 
or the functional integral formalism. 

In a coordinate representation, the stochastic NSE is written in the form 
\begin{equation}
\frac{\d \vu}{\d t}+ (\vu\cdot\nabla) \vu = \nu \Delta \vu
                    -\nabla\left(\frac{p}{\rho}\right) + \veta (\vx,t).
\label{snse}
\end{equation}
The random force correlator 
$\bra \eta_i(x) \eta_j(x') \ket = D_{ij}(x-x')$ 
should obey certain conditions to make the resulting theory physically 
feasible and the perturbation  expansion suitable for  analytical evaluation. 
First, the 
energy injection by random force should be equal to the energy 
dissipation; secondly, the forcing should be essentially 
infrared (IR), \ie it should be localized at large scales; third, it is desirable to 
have a parameter to control the IR divergences (when the size 
of the system tends to infinity). 

To exclude the pressure term in \eqref{snse} using the incompressibility 
condition, the Fourier representation is used     
\begin{equation}
\left(-\imath\omega + \nu\vk^2 \right)\hat u_i(\vk,\omega) - 
\int \dk{q}{d+1} M_{ijk}(\vk) \hat u_j(q)\hat u_k(k-q)=\hat\eta_i(k).
\label{feq2}
\end{equation} 
The random force correlator is usually taken in the form 
\begin{equation}
\bra \hat\eta_i(k_1)\hat\eta_j(k_2)\ket = (2\pi)^{d+1} \delta^{d+1}(k_1+k_2) 
          P_{ij}(\vk_1)D(\vk_1),
\label{noise-f}
\end{equation}
where the function $D(|\vk|)$ has a suitable power-law behavior. 
In the simplest,
but not very feasible physically, case $D(\vk)=D_0=const$, we deal with 
the white noise, $\delta$-correlated in both space and time. 

What is most realistic physically, is to have a random force concentrated  
in a limited domain in $k$-space $\Lambda_{min}\!<\!|\vk|\!<\!\Lambda_{max}$. 
This case, however, is difficult to evaluate analytically in the perturbation 
theory \cite{FNS1977}. In the multiscale approach we are going to present, 
we solve this problem 
by constructing a noise acting in a limited domain of 
scales $a$ in $(a,\vk)$ space. 

In the $(a,\vk)$ representation, having excluded the pressure by standard means of the orthogonal projector, the Eq. \eqref{snse} leads to a system 
of integro-differential equations for the scale components $\hat u_{ai}(k)$:
\begin{eqnarray}
(-\imath\omega+\nu\vk^2)\hat u_{ai}(k)&=& \hat\eta_{ai}(k)\label{fak1} \\
\nonumber                             &+& \left(\frac{2}{C_\psi}\right)^2
\int 
M_{ijk}^{aa_1a_2}(\vk,\vq,\vk\!-\!\vq)\hat u_{a_1j}(q)\hat u_{a_2k}(k\!-\!q)
\frac{da_1}{a_1}\frac{da_2}{a_2}\frac{d^{d+1}q}{(2\pi)^{d+1}}.
\end{eqnarray}
Now we face the problem of appropriate choice of the force correlator 
$$\bra \hat\eta_{ia_1}(k_1) \hat\eta_{ja_2}(k_1) \ket = 
D^{a_1a_2}_{ij}(k_1,k_2).$$
It was shown in the previous paper \cite{AltDAN03e} that the $(a,k)$ 
representation 
provides an extra analytical flexibility in constructing random processes 
with desired correlation properties in a coordinate space. For instance, 
the random process given by wavelet coefficients with the correlation 
function 
\begin{equation}
\bra\hat\eta_{a_1}(\vk_1)\hat\eta_{a_2}(\vk_2)  \ket = 
(2\pi)^d \frac{C_\psi}{2} \delta^d(\vk_1+\vk_2) a_1 \delta(a_1-a_2) D_0
\label{white-a}
\end{equation}
possesses the same correlation properties in $\R^d$ coordinate space 
as the white noise does. 
Casting $\hat\eta(k)$ in terms of $\hat\eta_a(k)$ by means 
of \eqref{wtf} we get 
\begin{eqnarray*}
\bra\hat\eta(\vk_1)\hat\eta(\vk_2)  \ket &=& \left(\frac{2}{C_\psi}\right)^2 
\int \frac{da_1}{a_1}\frac{da_2}{a_2} \hat\psi(a_1\vk_1)\hat\psi(a_2\vk_2) 
\bra\hat\eta_{a_1}(\vk_1)\hat\eta_{a_2}(\vk_2)  \ket \\ 
&=& (2\pi)^d \frac{2D_0}{C_\psi} \delta^d(\vk_1+\vk_2) 
\int \frac{da_1}{a_1}\frac{da_2}{a_2} \hat\psi(a_1\vk_1)\hat\psi(a_2\vk_2) 
a_1 \delta(a_1-a_2)\\
 &=&  (2\pi)^d \delta^d(\vk_1+\vk_2) D_0  
\end{eqnarray*}
that coincides with the correlation function of the white noise. 
The direct wavelet transform of the white noise 
$\eta(\vx)\!\to\!\hat\eta(\vk)\!\to\!\hat\eta_a(\vk)$ 
apparently leads to another result 
\begin{equation}
\bra\hat\eta_{a_1}(\vk_1)\hat\eta_{a_2}(\vk_2)  \ket = 
(2\pi)^d \delta^d(\vk_1+\vk_2) D_0 
\overline{\hat\psi(a_1\vk_1)\hat\psi(a_2\vk_2)}, 
\label{white-d}
\end{equation}
which is different from \eqref{white-a} and explicitly depends on the  
basic wavelet $\psi$. 

Physically, the  scale-dependent processes obeying \eqref{white-a} and \eqref{white-d}, 
respectively, describe quite different processes:  fluctuations of the former 
type \eqref{white-a} are mutually 
correlated only for coinciding scales ($a_1\!=\!a_2$), while for the latter 
case \eqref{white-d} all fluctuations are correlated.

Exactly as in the standard wavenumber space approach 
\cite{FNS1977,YO1986,AAV1999}, 
we can generalize the 
$\delta$-correlated force \eqref{white-a} assuming its variance  
to be dependent on both the scale the wave vector: $D_0 \to D(a,\vk)$. 
Taking into account we deal with 
the incompressible fluid in $d$ dimensions, we can put down a  
general form of the desired force correlator
\begin{equation}
\bra \hat \eta_{a_1i}(k_1) \hat \eta_{a_2j}(k_2) \ket = 
(2\pi)^{d+1} \delta^{d+1}(k_1+k_2) \frac{C_\psi}{2} a_1 \delta(a_1-a_2) P_{ij}(\vk_1) 
D(a_1,|\vk_1|).
\label{mstc} 
\end{equation}   
The $\delta$-correlated random force in the wavenumber space does not 
provide an adequate description of hydrodynamic turbulence, for it gives an 
energy injection in all scales, small and large. In physical settings, the 
fluid 
is usually stirred  at a predetermined scale, or in a narrow range 
of scales, comparable to the size of the system. 
As a simplest model of such a forcing, we can consider a force acting on 
a single scale $a_0$ by choosing 
\begin{equation}
D(a,\vk) = D_0 a_0 \delta(a-a_0).
\label{a0band}
\end{equation}  

Now let us turn to the perturbative calculations.
The stochastic diagram techniques for the component fields 
$\hat\vu_a(k)$ stems from Eq. \eqref{fak1} and is a straightforward 
generalization of the Wyld diagram technique for the Fourier components 
$\hat\vu(\vk)$:  
\begin{eqnarray}
\hat u_{ai}(k) &=& G_0(k) \hat\eta_{ai}(k) + G_0(k)  
 \left(\frac{2}{C_\psi}\right)^2 \int \frac{da_1}{a_1} \frac{da_2}{a_2}\dk{q}{d+1} 
\label{tree1} \\ 
\nonumber  & & M_{ijk}^{aa_1a_2}(\vk,\vq,\vk-\vq)  
\hat u_{a_1j}(q)  \hat u_{a_2k}(k-q), 
\end{eqnarray}
where $G_0(k) = (-\imath\omega + \nu\vk^2)^{-1}$ is the bare  
response function for Fourier component.
To keep the scales and wavevectors on the same footing and make the notation covariant 
in that sense, we can rewrite \eqref{tree1} using the response functions bearing 
scale indices explicitly \eqref{brf}. Thus, we get 
\begin{eqnarray}
\hat u_{ai}(k) &=& \int \frac{da_0}{a_0} G_{0ij}^{aa_0}(k) \hat\eta_{a_0j}(k) +   
 \left(\frac{2}{C_\psi}\right)^2 \int \frac{da_0}{a_0} G_{0il}^{aa_0}(k)
\frac{da_1}{a_1} \frac{da_2}{a_2}\dk{q}{d+1} 
\label{tree1a} \\ 
\nonumber  & & M_{ljk}^{aa_1a_2}(\vk,\vq,\vk-\vq)\hat u_{a_1j}(q)
\hat u_{a_2k}(k-q)    \label{iter0} \\
\nonumber G_{0ij}^{aa_0}(k) &=& \frac{\delta(a-a_0)a_0}{-\imath\omega+\nu\vk^2} 
\delta_{ij}.  
\end{eqnarray}
The Feynman expansion for the scale component fields $\hat u_{ai}(k)$ 
can be derived either from \eqref{tree1} or \eqref{tree1a}.  
Iterating the Eq. \eqref{tree1} once, we get the one-loop contribution 
to the response function: 
\begin{eqnarray}
\hat u_{ai}(k) &=& G_0(k) \hat\eta_{ai}(k) + G_0(k)  
 \left(\frac{2}{C_\psi}\right)^2 
\int \frac{da_1}{a_1} \frac{da_2}{a_2}\dk{k_1}{d+1} 
\label{iter1} \\ 
\nonumber  & & M_{ijk}^{aa_1a_2}(\vk,\vk_1,\vk-\vk_1)  \hat u_{a_1j}(k_1) 
\Bigl[G_0(k\!-\!k_1) \hat\eta_{a_2k}(k\!-\!k_1) \\ 
\nonumber &+& G_0(k\!-\!k_1)\left(\frac{2}{C_\psi}\right)^2
\int\frac{da_3}{a_3}\frac{da_4}{a_4}\dk{k_2}{d+1}
M_{klm}^{a_2a_3a_4}(\vk\!-\!\vk_1,\vk_2,\vk\!-\!\vk_1\!-\!\vk_2)  \\
\nonumber  & & \hat u_{a_3l}(k_2)\hat u_{a_4m}(k\!-\!k_1\!-\!k_2)\Bigr].
\end{eqnarray}
As usual, we assume the random force to be gaussian, with all odd correlators $\bra\eta_1\ldots \eta_{2k+1}\ket$ vanish identically. 

Following \cite{FNS1977}, we introduce a formal parameter of the 
perturbation expansion $\lambda$ in the interaction term 
($M^{aa_1a_2}_{ijk}\!\to\!\lambda M^{aa_1a_2}_{ijk}$).  
In  the final end the 
initial value $\lambda\!=\!1$ should be restored. The validity of considering 
$\lambda$ as a small parameter of perturbation expansion is justified 
by renormalization group methods \cite{FNS1977,YO1986,AAV1999}; see also 
\cite{AHHJ2003} for recent developments and generalizations. 

In the zero-th order of perturbation expansion the response function 
does not depend on 
scale and coincides with $G_0$ for all scale components:
\begin{equation}
\hat u_{ai}(k) = G_0(k) \hat\eta_{ai}(k).
\label{order0}
\end{equation}
In the $O(\lambda^2)$ and the next orders of perturbation expansion the 
standard stochastic diagram techniques, used by many authors 
\cite{Wyld1961,FNS1977,AAV1999}, is reproduced with the difference that: 
(i) each vertex, each response and correlation function attain scale superscripts; 
(ii) integration over octaves $\int\frac{da}{a}$ is performed over all 
pair-matching  
scale indices. Mathematically, this means that each diagram line corresponding to 
the plane wave component $\hat u_i(k)$ in standard techniques now attains an 
extra wavelet factor and becomes 
$\hat u_{ai}(k)=\overline{\hat\psi(a\vk)}\hat u_i(k)$. The Feynman graphs and 
their  topological factors of course remain the same. 

\begin{figure}
\centering \includegraphics[width=12cm]{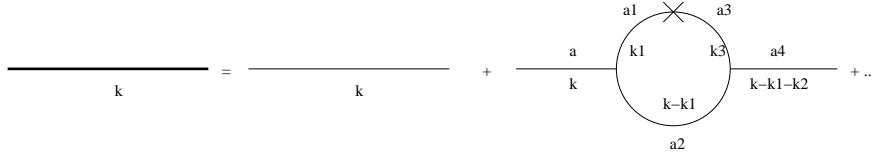}
\caption{One loop contribution to the response function}
\label{resp1l:pic}
\end{figure}

\begin{figure}
\centering \includegraphics[width=12cm]{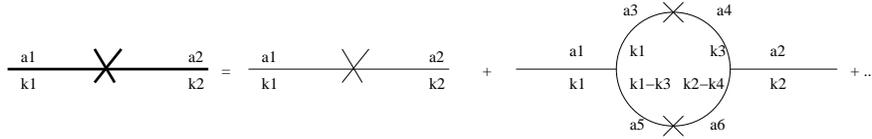}
\caption{One loop contribution to the correlation function}
\label{cor1l:pic}
\end{figure}

Here, for bookkeeping reasons, we present only one loop 1PI contributions 
to the response, see Fig.~\ref{resp1l:pic}, and correlation, 
see Fig.~\ref{cor1l:pic}, functions of stochastic hydrodynamics. 
In the first order in the force correlator $\bra\eta\eta\ket$ (one-loop 
contribution) we 
substitute the scale components $\hat u$ in the r.h.s. of \eqref{iter1} by 
the zero-th order solutions \eqref{order0}, perform necessary scale 
integrations in $a_l \delta(a-a_l) \frac{da_l}{a_l}$ and average over 
the random force \eqref{mstc}: 
\begin{eqnarray*}
 \hat u_{ai}(k) &=& G_0(k) \hat\eta_{ai}(k) + G_0(k)  
 4\lambda^2\left(\frac{2}{C_\psi}\right)^4
\int \frac{da_1}{a_1} \frac{da_2}{a_2} \frac{da_3}{a_3} \frac{da_4}{a_4}
\dk{k_1}{d+1} \dk{k_2}{d+1}
 \\ 
  & & M_{ijk}^{aa_1a_2}(\vk,\vk_1,\vk\!-\!\vk_1)  G_0(k_1)
\bra \hat\eta_{a_1j}(k_1)\hat\eta_{a_3l}(k_2) \ket G_0(k_2) G_0(k\!-\!k_1)  \\
 & & M_{klm}^{a_2a_3a_4}(\vk-\vk_1,\vk_2,\vk-\vk_1-\vk_2) 
  G_0(k\!-\!k_1-\!k_2)\hat\eta_{a_4m}(k-k_1-k_2).
\end{eqnarray*}
The corresponding diagram is shown in Fig.~\ref{resp1l:pic}.
The factor 4 accounts for two possible ways to expand 
the nonlinear term multiplied by two possible ways of taking 
the random force averaging. 
After substituting the random force correlator \eqref{mstc},  
the using of the explicit form  of the interaction 
vertexes \eqref{wak}, and integrating over the scales, this leads to   
\begin{eqnarray}
 \hat u_{ai}(k) &=&  G_0(k) \hat\eta_{ai}(k) 
+ \overline{\hat\psi(a\vk)}G_0^2(k)  
 4\lambda^2 \left(\frac{2}{C_\psi}\right)\int \dk{k_1}{d+1}  \frac{da_4}{a_4} 
M_{ijk}(\vk)  \label{G2eq} \\
\nonumber & &|G_0(k_1)|^2 \Delta_{jl}(\vk_1) G_0(k\!-\!k_1)  
M_{klm}(\vk-\vk_1)   \hat\psi(a_4\vk)\hat\eta_{a_4m}(k), \\
\Delta_{jl}(\vk_1) &=& P_{jl}(\vk_1)\left(\frac{2}{C_\psi}\right) 
\int \frac{da_1}{a_1} |\hat\psi(a_1\vk_1)|^2 D(a_1,\vk_1). \label{effcor}
\end{eqnarray}
Eq.\eqref{effcor} means that for our special 
type of the scale-dependent random 
forcing \eqref{mstc} all internal parts of the diagrams, that 
do not carry the scale indices, can be evaluated by substitution of the effective 
force correlator \eqref{effcor} into the standard diagrams drawn in wavenumber space.
In this way we can easily evaluate the perturbative corrections to the usual 
response function $G(k)$ for Fourier components (see Appendix), and hence 
evaluate the turbulent corrections to viscosity, produced by 
scale-dependent force.

Similarly, we can evaluate the contributions 
to the correlation functions $\bra\hat u_{a_1i}(k_1)\hat u_{a_2j}(k_2)\ket$.
In the one-loop approximation, using \eqref{iter1} and the zero-th order 
approximation \eqref{order0}, we get 
\begin{eqnarray*}
\bra\hat u_{a_1i}(k_1)\hat u_{a_2j}(k_2)\ket &=& G_0(k_1)G_0(k_2)
\bra\hat \eta_{a_1i}(k_1)\hat \eta_{a_2j}(k_2)\ket \\
&+& 2 \lambda^2 G_0(k_1)G_0(k_2) \left(\frac{2}{C_\psi}\right)^4 
\int \frac{da_3}{a_3}\frac{da_5}{a_5}
\frac{da_4}{a_4}\frac{da_6}{a_6}\dk{k_3}{d+1}\dk{k_4}{d+1} \\
& & M_{ilk}^{a_1a_3a_5}(\vk_1,\vk_3,\vk_1\!-\!\vk_3)
G_0(k_3)\bra\hat \eta_{a_3l}(k_3)\hat \eta_{a_4m}(k_4)\ket G_0(k_4) 
G_0(k_1\!-\!k_3)\\
& & \bra\hat \eta_{a_5k}(k_1\!-\!k_3)\hat \eta_{a_6n}(k_2\!-\!k_4)\ket 
G_0(k_2\!-\!k_4) M_{jmn}^{a_2a_4a_6}(\vk_2,\vk_4,\vk_2\!-\!\vk_4).
\end{eqnarray*}  
After the integrations over $a_4,a_6,k_4$, identically to the response 
functions calculations, with the random force \eqref{mstc},
 we get the one-loop contribution to the correlation function, shown 
in Fig.~\ref{cor1l:pic}: 
\begin{eqnarray}
\nonumber \bra\hat u_{a_1i}(k_1)\hat u_{a_2j}(k_2)\ket &=& |G_0(k_1)|^2
\bra\hat \eta_{a_1i}(k_1)\hat \eta_{a_2j}(k_2)\ket 
+ 2\lambda^2 \delta^{d+1}(k_1+k_2)\left(\frac{2}{C_\psi}\right)^2 
\int \frac{da_3}{a_3}\frac{da_5}{a_5}
\dk{k_3}{d+1} \\
\nonumber & & M_{ilk}^{a_1a_3a_5}(\vk_1,\vk_3,\vk_1\!-\!\vk_3)
|G_0(k_3)|^2P_{lm}(\vk_3)D(a_3,\vk_3)|G_0(k_1\!-\!k_3)|^2\\
     & & P_{kn}(\vk_1\!-\!\vk_3)D(a_5,\vk_1\!-\!\vk_3) 
M_{jmn}^{a_2a_4a_6}(-\vk_1,-\vk_3,-\vk_1\!+\!\vk_3).\label{l1c}
\end{eqnarray}
Expanding the wavelet factors $\psi(a\vk)$ in each vertex and 
integrating 
over all matching scale arguments, we get the one-loop correction to the 
correlation function 
\begin{eqnarray}
\nonumber C_2(a_1,k_1,a_2,k_2) &=& 2 \delta^{d+1}(k_1\!+\!k_2)\lambda^2|G_0(k_1)|^2 
\overline{\hat\psi(a_1\vk_1)\hat\psi(-a_2\vk_1)} \int \dk{k_3}{d+1}
M_{ilk}(\vk_1) \\
& & |G_0(k_3)|^2\Delta_{lm}(\vk_3) |G_0(k_1\!-\!k_3)|^2\Delta_{kn}(\vk_1\!-\!\vk_3)
M_{jmn}(-\vk_1).\label{c2k}
\end{eqnarray} 
The evaluation of the one-loop contribution to the correlation function 
is easily performed for the above mentioned narrow-band force correlators. 
The contribution 
takes the form 
$$C_2(a_1,k_1,a_2,k_2) = \delta^{d+1}(k_1+k_2)
\hat\psi(a_1\vk_1)\psi(-a_2\vk_2)C_{eff}(k_1)$$
\begin{equation}
C_{eff}(k) = 2 \lambda^2|G_0(k)|^2 \int \dk{q}{d+1} \Delta(\vq)\Delta(\vk-\vq)
c_2(\vk,\vq) |G_0(q)|^2 |G_0(k-q)|^2,
\label{c2eff}
\end{equation}
where the trace of the one-loop tensor structure $c_2(\vk,\vq)$ is given 
in Appendix. 
The integration in frequency argument in the limit of zero frequency 
($k_0\!\to\!0$), is not different from that in stochastic hydrodynamics 
in the wavenumber space and gives 
\begin{eqnarray}
\nonumber C_{eff}(\vk) &=& \lambda^2|G_0(0,\vk)|^2 S_{d-1}\int 
\frac{
q^{d-1}dq d\theta \sin^{d-2}\theta\Delta(\vq)\Delta(\vk-\vq)
}{
\nu^3 q^2(k^2 -2 k q\cos\theta +q^2)(k^2 -2 k q\cos\theta +2q^2)} \\
& & \frac{(1-\cos^2\theta) k^2}{4(k^2-2kq\cos\theta +q^2)}
\bigl[k^2(d-1) - 2kqd\cos\theta +2q^2(d+2\cos^2\theta-2) \bigr].
\label{c2e1}
\end{eqnarray}
As an example, let us present the one-loop contribution to the effective 
pair correlator \eqref{c2eff}, calculated for the case of single-scale 
forcing \eqref{a0band} with the basic wavelets from the gaussian vanishing 
momenta wavelet family 
\begin{equation}
\hat g_n(k) = (2\pi)^\frac{d}{2}(-\imath \vk)^n \exp(-\vk^2/2), \quad 
C_{g_n} = (2\pi)^d \Gamma(n),
\label{vmf}
\end{equation}
with $\Gamma(x)$ being the Eulerian gamma-function. 

For the single-scale forcing \eqref{a0band} this gives an effective 
force correlator in the wavenumber space
\begin{equation}
\Delta_n(q) = \frac{D_0}{\Gamma(n)} (a_0q)^{2n}e^{-(a_0q)^2}.
\label{effn}
\end{equation}
Straightforward calculation leads to: 
\begin{eqnarray}
 C_{eff}(k) &=& \lambda^2
\frac{k^2 |G_0(k)|^2 S_{d-1}a_0^{4n}D_0^2}{4\nu^3\Gamma(n)^2} 
\int 
\frac{
q^{2n}(k^2-2kq\cos\theta+q^2)^{n}e^{-a_0^2(k^2-2kq\cos\theta+2q^2)}
}{q^2(k^2-2kq\cos\theta+q^2)^2 (k^2-2kq\cos\theta+2q^2)} \label{cefn}\\ 
\nonumber & &(1-\cos^2\theta)
\bigl[k^2(d-1)-2kqd\cos\theta+2q^2(d+2\cos^2\theta-2) \bigr]
q^{d-1}dq \sin^{d-2}\theta d\theta.
\end{eqnarray}
The integration over the angle variable $\cos\theta$ can be performed 
explicitly. 
With the calculations presented in Appendix, 
we get the effective pair correlator for $n=2,d=3$ in the large-scale limit 
$\left(x=\frac{k}{q}\!\to\!0\right)$:
\begin{equation}
C_{eff}^{d=3,n=2}(k\to0) = \frac{7}{40}
\frac{k^2 |G_0(k)|^2 \pi^\frac{3}{2}a_0^3 D_0^2}{\nu^3\sqrt{2}}\lambda^2. 
\end{equation}

\section{Energy dissipation and energy transfer}
The energy dissipation rate per unit of mass of an incompressible viscous 
fluid 
is given by the Navier-Stokes equations 
$$
\epsilon = -\nu \int u(x) \Delta u(x) d^d\vx = \nu \int d^d\vx (\nabla u)^2.
$$
Using wavelet decomposition \eqref{icwt} for velocity field we get 
\begin{equation}
\epsilon = -\nu \int \Omega(a_1,a_2,\vb_1-\vb_2)  u_{a_1}(b_1) u_{a_2}(b_2) 
\frac{da_1d^d\vb_1}{a_1}\frac{da_2d^d\vb_2}{a_2},
\label{edis1}
\end{equation}
where
\begin{eqnarray*}
\nonumber \Omega(a_1,a_2,\vb_1-\vb_2) &=& 
\left(\frac{2}{C_\psi}\right)^2 \int \frac{d^d\vx}{(a_1a_2)^d} 
\psi\left(\frac{\vx-\vb_1}{a_1}\right)\frac{\d^2}{\d \vx^2}
\psi\left(\frac{\vx-\vb_2}{a_2}\right)  \\
&=& -\left(\frac{2}{C_\psi}\right)^2 \int \vk^2 
\hat\psi(a_1\vk)\hat\psi(-a_2\vk)e^{-\imath\vk(\vb_1-\vb_2)}\dk{\vk}{d} 
\label{lapcon}  
\end{eqnarray*}
is the dissipation connection for the scale components. 

In symbolic form,  
the contribution of the fluctuations of all scales $a_i$ to the mean dissipation 
of energy per unit of mass can be written as 
\begin{equation}
\epsilon = \sum_{ij} \nu_{ij} \int u_{a_i}(x_i) u_{a_j}(x_j) d^d\vx_id^d\vx_j, 
\end{equation}
where $\nu_{ij}$ is the viscosity between $a_i$ and $a_j$ scales. 
Being well localized in both the real and the wavenumber space, the analyzing 
function $\psi$ perceives  
the interaction of the components of the same or close scales stronger  
than the contributions of the significantly different ones 
$|\log(a_1/a_2)|\gg1$. For the Daubechies wavelets (the orthogonal 
wavelets with compact support) often used for numerical simulation of 
turbulence, the viscosity connection coefficients can be found elsewhere 
\cite{LRT1991}. 

For a qualitative estimation of the behavior of the viscosity connection 
as a function of the scale ratio $a_1/a_2$ of the interacting scale components, 
let us consider the vanishing momenta wavelet family of gaussian wavelets 
\eqref{vmf}, considered by Lewalle in a wavelet-based analysis of energy 
dissipation \cite{Lewalle1993pf,Lewalle1993as}. 
The viscosity connection 
\eqref{lapcon} can be then evaluated analytically:
\begin{eqnarray}
\nonumber \Omega_n(a_1,a_2,b_1-b_2) &=& -\frac{(a_1 a_2)^{n}}{\Gamma^2(n)} 
\int_{-\infty}^{\infty} 
(k^2)^{n+1} e^{-\frac{k^2}{2}(a_1^2+a_2^2) -\imath (b_1-b_2)k} 
\dk{k}{d} \\
&=& \frac{(a_1a_2)^n}{\Gamma^2(n)}(-1)^n 2^{n+1}\left.{ 
\frac{d^{n+1}}{d\sigma^{n+1}} 
\frac{e^{-\frac{(b_1-b_2)^2}{2\sigma}}}{\sigma^{d/2}}
}\right|_{\sigma = a_1^2+a_2^2}. 
\end{eqnarray}

\begin{figure}[h]
\centering \includegraphics[angle=270,width=8cm]{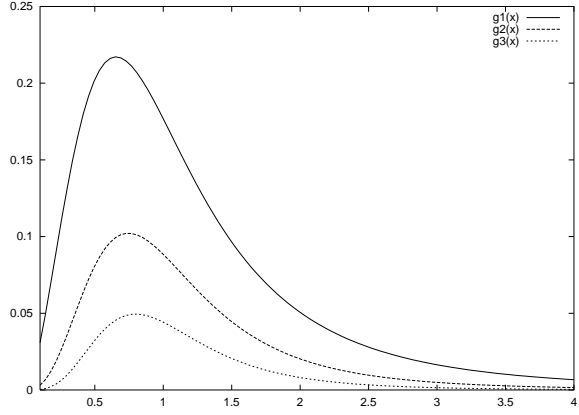}
\caption{Graph of the function $\frac{1}{a_2^2(1+t^2)} 
\left({\frac{t}{t^2+1}}\right)^\frac{2n+1}{2}$ for $n=1,2,3$ related to the 
viscosity connection for gaussian type wavelets.}
\label{vmfcon:pic}
\end{figure}

The main contribution to energy dissipation comes from the terms with 
coinciding or closed arguments $x=b_1-b_2\approx0$. 
In this limit (with $d=1$, taken for simplicity) we get 
$$\Omega_n \sim \frac{1}{a_1^2+a_2^2} 
\left(\frac{a_1a_2}{a_1^2+a_2^2}\right)^\frac{2n+1}{2} (2n+1)!!.$$
Introducing the ratio  $t=\frac{a_1}{a_2}$, we can study the behavior of 
the viscosity connection as a function of scale ratio  
$$\Omega_n \sim  \frac{1}{a_2^2(1+t^2)} 
\left(\frac{t}{t^2+1}\right)^\frac{2n+1}{2}
(2n+1)!!.$$
The plot of the viscosity connection $\Omega_n(a_1,a_2,0)$ 
as a function of the scale ratio $t=\frac{a_1}{a_2}$, for the first three  
wavelets ($n=1,2,3$) of the \eqref{vmf} family,  
is shown in Fig.~\ref{vmfcon:pic}. As it can be seen, regardless the number of 
vanishing momenta $n$, the dissipation 
term has a maximum at coinciding scales ($t\approx1$).  
For this reason, if a discrete 
wavelet decomposition  is used instead of continuous one, 
it is sufficient to keep the two 
main terms in the energy dissipation: the equal-scale interaction 
and the neighboring scale interaction
\begin{equation}
\epsilon \sim -u^j_k u^j_m \int dx \psi^j_k(x) \Delta \psi^j_m(x) 
- 2 u^{j-1}_k u^j_m \int dx \psi^{j-1}_k(x) \Delta \psi^j_m(x) + \ldots.
\label{dis2}
\end{equation}
The first term is the standard viscosity term, the last is the Kraichnan 
nearest scales interaction. 

The nonlinear energy transfer between neighboring scales can be evaluated by 
considering the wavelet connections corresponding to nonlinear term 
$(u\nabla)u$ of the Navier-Stokes equation. To keep with the wavelet 
turbulence cascade models \cite{Zimin1981}, we restrict ourselves with a discrete wavelet transform 
with binary scale step, $a_0=\frac{1}{2}$: 
\begin{equation}
\vu(x) = \sum_{j\vk} \vu_\vk^j \psi_\vk^j(\vx) + \hbox{Error term}, 
\quad \psi_\vk^j(\vx) \equiv a_0^{-\frac{jd}{2}} 
\psi \left(\frac{\vx-\vk b_0 a_0^j}{b_0 a_0^j} \right)
\label{dwt}.
\end{equation}
Without loss of generality the mesh size is set to unity $b_0\!=\!1$. 
In our consideration of the DWT representation \eqref{dwt} applied to 
hydrodynamic turbulence, in contrast to many schemes applied for numerical 
simulation of the NSE \cite{QW1993,Reslea1995}, 
we have no {\em a priory} 
arguments to assume a mutual orthogonality of the basis functions $\psi_k^j$. 
To keep the wavelet decomposition \eqref{dwt} unique, the 
orthogonality of basic functions is not required. It is sufficient if 
the set of basic functions forms a {\em frame}, \ie $\forall f \in \lr, 
\exists A>0, B<\infty$ such that 
$$
A\|f\|^2\le \sum_{jk} |\bra f|\psi_k^j\ket|^2 \le B \|f\|^2.
$$    
If $A=B$, the frame is called a {\em tight frame}.

Assuming the basic functions $\psi^j_k$ form a frame, and restricting  
ourselves with the case of incompressible fluid, we cast the NSE system 
in the form 
\begin{equation}
\frac{\d u_k^{j\alpha}}{\d t} \psi^j_k(x) + u_m^{l\beta}\psi_m^l(x) 
\frac{\d u_k^{j\alpha}\psi_k^j(x)}{\d x^\beta} = 
-\frac{1}{\rho}\frac{\d p_k^j \psi_k^j(x)}{\d x^\alpha} 
+\nu u_k^{j\alpha}\Delta \psi_k^j(x),
\label{dwtnse}
\end{equation} 
with Greek letters used for the coordinate indices, the bold face for 
the vector subscripts being dropped, and the summation over all pair-matching 
indices assumed. The component fields 
$u_k^{j\alpha} =u_k^{j\alpha}(t)$ are the functions of time only;  
so we deal with a typical cascade model. 

Our goal at this point is to derive the 
energy transfer between the components of $j$-th scale and the next small 
$(j\!+\!1)$-th one. 
Let us define the energy of the $j$-th scale pulsations as 
\begin{equation}
E_j = \frac{1}{2} \sum_{\alpha k} 
\bar u_k^{j\alpha}u_m^{j\alpha}\Lambda_{km}^j, 
\quad \Lambda_{km}^j = \int d^d x \bar \psi_k^j(x)\psi_m^j(x), 
\end{equation}      
with the assumed unit normalization of the basic function 
$\int d^dx |\psi(x)|^2 = 1$. 
The contribution of the nonlinear term of the NSE to the time derivative 
of $E_j$ is
\begin{equation}
\Delta E_j = -\Delta t \bar u_s^{r\alpha} u_m^{l\beta}  u_k^{j\alpha}
\int d^d x \bar \psi_s^r(x) \psi_m^l(x) \frac{\d \psi^j_k(x)}{\d x^\beta}.
\label{det}
\end{equation}
For the orthogonal wavelets, that are most often used in numerical 
simulations \cite{QW1993}, only the terms of coinciding scales 
$r=l=j$ survive in the r.h.s. of \eqref{det}. The energy flux from 
the $j$-th scale to the next $(j\!+\!1)$-th scale is then proportional 
to $|u^j|^3/(b_0 a_0^j)$, in exact accordance to the Kolmogorov 
phenomenological theory \cite{K41a}. In more general terms of nonorthogonal 
basic functions, the next term in the r.h.s. of \eqref{det} is 
proportional to $u^{j+1} u^{j+1} u^j$. This term can be interpreted as 
the material derivative $u^j \nabla (u^{j+1})^2$ of the mean energy  
$\frac{(u^{j+1})^2}{2}$ of the 
small scale fluctuations travelling along the stream of 
large scale velocity $u^j$.

The energy transfer terms analogues to \eqref{det} have been already 
considered in the orthogonal wavelet formalism by C.Meneveau \cite{Meneveau1991}. 
They can be obtained directly from the component equations, by multiplying 
\eqref{fak1} by $\overline{\hat u_{ai}}(k)$. This leads to the energy 
transfer in $(a,\vk)$ space 
\begin{equation}
t(a,k)= \left(\frac{2}{C_\psi}\right)^2 \int \overline{\hat u_{ai}}(k) 
M_{ijk}^{aa_1a_2}(\vk,\vq,\vk\!-\!\vq)\hat u_{a_1j}(q)\hat u_{a_2k}(k\!-\!q)
\frac{da_1}{a_1}\frac{da_2}{a_2}\frac{d^{d+1}q}{(2\pi)^{d+1}},
\label{fak2}
\end{equation} 
given in \cite{Meneveau1991}.

\section{Kolmogorov hypotheses}
The Kolmogorov theory of the locally isotropic turbulence is formulated in 
terms of relative velocities
\begin{equation}
\delta u(r,l) = u(r+l) - u(r).
\label{incr}
\end{equation}
The probability distribution of relative velocities \eqref{incr} 
hardly can be  studied by the   
Fourier transform, in case the velocity field $u(r)$  is not homogeneous. 
According to Kolmogorov \cite{K41a}, the turbulence in space-time domain 
$G$, is referred to as a stationary turbulence  if, for any fixed $u(r,t)$, the distribution of relative velocities $\delta u(r,t)$ is stationary and isotropic. 
Physical assumptions on the locally isotropic turbulence were formulated in terms 
of the first and second Kolmogorov hypotheses, that reside on the definition of 
the Reynolds number. This is not a quite rigorous mathematical definition.  
We shall show  that the Kolmogorov hypotheses are the statements 
about the behavior of the scale components of velocity field.  

First, we have to note that the definition of the Reynolds number is consistent 
within the multiscale framework. In fact, by definition 
\begin{equation}
Re_l = \frac{u_l l}{\nu},
\label{Re}
\end{equation}
where $u_l$ are said to be ``pulsations of the scale $l$'', the rigorous 
definition of those can be given by virtue of wavelet components 
\eqref{dcwt}, 
considering $\psi$ as an apparatus function used to measure the pulsations. 
Going 
further, we find the wavelet components \eqref{dcwt} to be identical to 
velocity increments \eqref{incr}, in case $\psi$ is the Haar wavelet:
\begin{equation}
h(x) = \begin{cases}
       1 & 0\le x < 1/2 \\
       -1 & 1/2 \le x < 1 \\
       0  & \hbox{otherwise}
       \end{cases}.
\label{haar}
\end{equation}
Now let us consider the Kolmogorov hypotheses \cite{K41a}: 
\paragraph{H1: The first  hypothesis of similarity}
For the locally isotropic turbulence with high enough Re the PDFs for 
the relative velocities \eqref{incr} 
are uniquely determined by the viscosity $\nu$ and the mean energy dissipation 
rate $\epsilon$.
\paragraph{H2: The second hypothesis of similarity}
Under the same assumption as for H1 the turbulent flow is self-similar 
in small (but still $l\gg \nu^{\frac{3}{4}}\epsilon^{-\frac{1}{4}}$) scales 
in the sense that 
\begin{equation}
\delta u(r,\lambda l) \stackrel{law}{=} \lambda^h \delta u(r,l), \quad 
\lambda \in \R_+. 
\label{H2}
\end{equation} 
Using the Taylor frozen flow hypothesis, and therefore considering 
one-dimensional pulsations $u(x)$, according to the definition 
of wavelet coefficients \eqref{dcwt}, with $\psi\!=\!h$ we get 
\begin{eqnarray*}
u_l(r) = \int_{-\infty}^\infty \frac{1}{l} 
h \left(\frac{x-r}{l} \right) u(x) dx 
= \int_0^1 dt h(t) u(lt+r) \\
= \int_0^{1/2} u(lt+r) dt - \int_{1/2}^1 u(lt+r)dt. 
\end{eqnarray*} 
For small values of $l$ we can approximate  
$$
u_l(r) \approx \frac{1}{2}u(r+\frac{1}{4}l) -\frac{1}{2}u(r+\frac{3}{4}l), 
$$
or, taking into account the statistical homogeneity of the flow, we get
\begin{equation}
u_l(r) \approx \frac{1}{2} u(r) - \frac{1}{2} u(r+\frac{l}{2}).
\end{equation}
So, the power-law behavior \eqref{H2}, viz 
$$u_l(r) = -\frac{1}{2}\delta u(r,\frac{l}{2}) \sim l^h,$$
is just a particular case of a local regularity of wavelet 
coefficients, with the Haar function \eqref{haar} being used as a 
basic wavelet. As it was shown in general settings \cite{Holsh1990},
the wavelet coefficients $W_\psi(a,x)[f]$ of a square-integrable function 
$f(x)$, 
which has the Lipshitz-H\"older exponent $h$ at the point $x\!=\!x_0$, behave as 
$|W_\psi(a,x)[f]| \sim a^h$ inside the cone $|x-x_0| < const$ for any 
admissible wavelet $\psi$ which satisfies  the regularity condition 
\begin{equation}
\int_{-\infty}^{\infty} dx (1+|x|)|\psi(x)|<\infty.
\label{wreg}
\end{equation}
The condition \eqref{wreg} is rather loose, and in physical settings 
one can always assume that it holds for any analyzing function used to measure 
the pulsations of scale $l$. So, the second Kolmogorov hypothesis can 
be formulated as follows:
\paragraph{H2: Generalized second Kolmogorov hypothesis of similarity}
Under the same assumption as for H1 the turbulent flow is self-similar 
in small (but still $l\gg \nu^{\frac{3}{4}}\epsilon^{-\frac{1}{4}}$) scales 
in the sense, that the pulsations of the turbulent velocity defined as 
$$u_l(b) = \int \frac{1}{l} \bar\psi\left(\frac{x-b}{l} \right) u(x) dx,$$
where $\psi(x)$ is any analyzing function satisfying the admissibility 
condition \eqref{adc1} and the regularity condition \eqref{wreg}, have 
the following power-law behavior 
\begin{equation}
|u_l(b)|^2 \stackrel{law}{=} l^{2h} , \quad 
h = \frac{1}{3},
\label{H2g}
\end{equation} 
for all spatial points  $b$ occupied by turbulent media.

\section{Concluding remarks}
In this paper we have a stochastic hydrodynamics approach to the NSE based on 
a wavelet decomposition of both the velocity field and the stirring force used 
to balance the energy dissipation. In spite of a good deal of papers 
devoted do different choices of the stirring force for the NSE in wavenumber 
space -- see \eg. \cite{FNS1977,EP1988,AAV1999} and references therein -- 
what is essentially new in our approach is the definition of random 
force by the correlation function of its wavelet coefficients. Establishing 
the force correlator in the space of wavelet coefficients, we have got it 
easy to get rid of loop divergences in the stochastic perturbation 
expansion, having at the same time desired physical properties of the 
forcing (energy injection at a given scale). Therefore, a new UV-finite 
framework is constructed for statistical hydrodynamics. 
This is a technical framework for analytical evaluation of the statistical
characteristics of  turbulent fluctuations, such as their correlation and 
response functions, by means of continuous wavelet transform. 
Deriving physical 
consequences, such as energy cascade between scales or scale-dependent corrections 
to response function, we specially did not touch the renormalization group 
aspects \cite{FNS1977,FF1983} of the problem and multifractal formalism 
\cite{Benzi,AGH,ABM1991}.
In fact, both are related. The former is a generalization of the description 
of hydrodynamic turbulence in terms of differential equations to the  
description in measure settings. Partially, the relation of wavelets and RG in turbulence 
description is discussed in \cite{ffp5}, and will be studied in more details 
in connection to multifractal properties of hydrodynamic turbulence. Besides, 
the possible comparison with the RG based classification 
of asymptotic regimes of isotropic turbulence \cite{AAV1999,AHHJ2003} can 
be considered as another perspective of the proposed method.
  
The study of hydrodynamic turbulence by methods of stochastic differential 
equations and those of quantum field theory has at least half a century 
history.
Regardless phenomenologically clear and widely accepted Kolmogorov \cite{K41a} 
theory 
of fully developed turbulence, still there are discussions on the preference of 
either differential equations, or field theoretic methods based on renormalization 
group, or multifractal approach to describe the turbulence in an incompressible 
fluid. 

As it concerns the physical interpretation of the velocity field wavelet 
coefficients, by this paper we intended to say that stochastic 
nature of spatially 
extended hydrodynamic turbulence prescribes a certain kind of ``wave-particle 
dualism'' to the turbulent phenomena, in a sense, that the answer we get depends 
on the basic functions used to describe the turbulence. If the basis of plain 
waves was chosen, there are no fair reasons to comply about $k\!\to\!0$ 
behavior or paradoxes: what we get is what we set. Alternatively, if we want to 
have an analytical description of the spatially extended turbulence compatible with 
the Kolmogorov phenomenology of local turbulent pulsations of given scales, we 
need to set a functional basis that respects the scale locally. This is the wavelet 
decomposition.  

The Fourier transform, being essentially nonlocal, apparently does not fit 
the  above mentioned 
requirements, but the windowed Fourier transform, or wave packet decomposition 
used by V.Zimin \cite{Zimin1981} and T.Nakano for the analysis of 
turbulence does, and possibly there 
is only a technical difference between our approach and that of Nakano
\cite{Nakano1988}. 
However, it is important to emphasize that the incorporation of the basic 
function $\psi$ into consideration makes us to admit that the definition of the 
local fluctuations of a given scale is not completely objective and depends on 
means of observation. 
%At best it demands us to know the ``shape function'' of the 
%measuring device. Alternatively we can try to rely on those results, that are 
%shape-independent.
As it was shown, the Kolmogorov hypotheses (K41) were easily rewritten in 
the wavelet framework for a multiscale description of turbulence.   
   
\section*{Acknowledgement}
The author is thankful to Drs. M.Hnatich, J.Honkonen, M.Jurcisin and O.Chkhetiani for 
useful discussions. 
This work is supported in part by Russian Foundation for Basic Research, 
project 03-01-00657.

\appendix
\section{Calculation of one-loop diagrams}
\subsection{Response function}
To calculate the one-loop diagram in the response function, we introduce 
the following tensor structure  
\begin{equation}
L(k,q,a,s) = -m(k,a,b,c) o(q,b,l) m(p,c,l,s), 
\end{equation}
where the summation over all dummy indices is assumed. The following  
notation for the orthogonal projector and the vertex 
\eqref{feq1} is used ($p=k-q$): 
$$
o(q,b,l) = \delta_{bl} - \frac{q_b q_l}{q^2}, \quad  
m(p,c,l,s) = \frac{1}{2}\bigl[ p_l o(p,c,s) + p_s o(p,c,l) \bigr] .
$$
After all convolutions in matching pairs of indices, 
substituting $\vk\cdot\vq = k q \mu$, 
where $\mu\equiv\cos\theta$ is the cosine of the angle between $k$ and $q$, we get 
\begin{eqnarray*}
\nonumber L(k,q,a,s) &=& 
          \delta_{as}\frac{k^2(\mu^2-1)}{4} 
 + k_a k_s \frac{2k^2\mu^2 + 2\mu k q (1-2\mu^2) + p^2(1-4\mu^2)}{4p^2} \\
\nonumber &+&k_a q_s\frac{-k^2\mu^2 + k q\mu(2\mu^2-1)+\mu^2 p^2}{2p^2}        
  +q_a k_s \frac{-2 k^3\mu + 2 k^2 q (2\mu^2-1) + 3 k\mu p^2}{4p^2 q} \\ 
&+&q_a q_s \frac{k^3 \mu + k^2 q (1-2\mu^2) - k\mu p^2}{2q p^2}. 
\end{eqnarray*}
After substitution $p^2 = k^2 - 2kq\mu + q^2$ in the numerators and 
some algebraic simplification 
\begin{equation}
\nonumber L(k,q,a,s) = \frac{1}{2}\bigl[T_1 k^2 \delta_{as} + T_2 k_a k_s 
    + T_3 k_a q_s + T_4 q_a k_s + T_5 q_a q_s\bigr]
\label{rfts}
\end{equation}
where 
\begin{eqnarray*}
T_1 &=& \frac{\cos^2\theta-1}{2}, \\
T_2 &=& \frac{k^2+q^2 - 2\cos^2\theta(k^2 +2q^2)+ 4kq\cos^3\theta}{2p^2}, \\
T_3 &=& \frac{q^2\cos^2\theta - k q\cos\theta}{p^2}, \\
T_4 &=& \frac{k^3\cos\theta -2k^2q\cos^2\theta +3kq^2\cos\theta-2k^2q}{2q p^2},\\
T_5 &=&  \frac{k^2-kq\cos\theta}{p^2}.
\end{eqnarray*}
To calculate the whole one-loop integral  contribution to  
the response function, the tensor structure $L(k,q,a,s)$ is 
multiplied by the integral over the frequency component 
\begin{equation}
\int_{-\infty}^\infty \frac{dq_0}{2\pi} |G_0(q)|^2 G_0(k-q) = 
\frac{1}{2\nu^2\vq^2} \frac{1}{\frac{k_0}{\imath\nu}+ \vq^2 + (\vk-\vq)^2}.
\label{gfw0}
\end{equation}
Using this structure we easily get the 1PI one-loop contribution 
to the response function
\begin{eqnarray}
\nonumber \hat u_{i}(k) &=&  G_0(k) \hat\eta_{i}(k) 
+ G_0^2(k)  
 4\lambda^2 \int q^{d-1} dq d\theta \sin^{d-2}\theta d\phi   
\frac{1}{2\nu^2\vq^2}
\frac{\Delta(\vq)}{\frac{k_0}{\imath\nu}+ \vq^2 + \vp^2} 
\\ \nonumber & & \frac{1}{2}\bigl[
T_1 k^2 \delta_{as} + T_2 k_a k_s + T_3 k_a q_s + T_4 q_a k_s + T_5 q_a q_s
      \bigr] \hat\eta_s(k). \label{l2int} 
\end{eqnarray}
In $d=3$ we always 
assume $\vk = k \ve_z$, with $\theta$ being the polar angle 
$\vk\cdot\vq = kq\cos\theta$ and $\phi$ being the azimuthal angle. Let 
us evaluate the integral \eqref{l2int} in $d=3$ 
\begin{eqnarray}
\nonumber \hat u_{i}(k) &=&  G_0(k) \hat\eta_{i}(k) 
+ \lambda^2 G_0^2(k)  
 S_{2}\int q^{2} dq d\theta \sin\theta    
\frac{1}{\nu^2\vq^2}
\frac{\Delta(\vq)}{\frac{k_0}{\imath\nu}+ \vq^2 + \vp^2} 
\\ \nonumber & & {\rm diag}(T_1k^2 +\frac{q^2}{2}\sin^2\theta T_5,
T_1k^2 +\frac{q^2}{2}\sin^2\theta T_5, 0) \Bigr] \hat\eta_i(k) \label{l2resp} 
\end{eqnarray}
Here ${\rm diag}()$ means diagonal matrix, where the first two terms give 
the transversal contribution to viscosity, and the last term, 
giving the longtitudal contribution, vanishes identically: 
$$k^2 (T_1 + T_2) + k q \cos\theta (T_3 + T_4) + q^2 \cos^2\theta T_5=0.$$ 
Finally, after integration in angle variable $\cos\theta$, we get 
\begin{equation}
\hat u_{i}(k) =  G_0(k) \hat\eta_{i}(k) 
+ \lambda^2 G_0^2(k)  
 \frac{S_{2}}{\nu^2} \int_0^\infty  dq \Delta(\vq) 
{\rm diag}(R(k/q),R(k/q),0)\hat\eta_{i}(k) 
\end{equation}
where 
\begin{eqnarray}
\nonumber R(x) &=& \frac{1}{32x^3} \Bigl[ 12x-16x^3-8x^5 
+(x^2-1)^3 \ln \left(\frac{x-1}{x+1} \right)^2 \\
&+& (3x^6 -2x^4 + 12x^2 -8) \ln \frac{2+2x+x^2}{2-2x+x^2}\Bigr].
\end{eqnarray}
\subsection{Correlation function}
Similarly, for the tensor structure of the one-loop contribution to 
the response function, we evaluate the tensor structure of the one-loop 
contribution to the pair correlator 
$$C(k,q,a,s) = m(k,a,b,c) m(k,s,l,t) o(q,b,l) o(p,c,t).$$
After algebraic simplification 
\begin{eqnarray}
\nonumber C(k,q,a,s) &=& 
 \delta_{as}
 \frac{-k^4 + 2k^3\mu q + k^2(-\mu p^2-\mu^2 q^2+2p^2)}{4p^2}   \\ 
\nonumber &+&  k_a k_s 
 \frac{k^2(1+\mu^2)-2k q\mu (1+2\mu^2) + 2(2\mu^4q^2-p^2)}{4p^2} \\
\nonumber &+&(k_a q_s + q_a k_s) 
\frac{-k^3\mu + 4k^2 q \mu^2 + k\mu (-4\mu^2 q^2 + p^2 + q^2)}{4p^2 q} \\ 
&+&q_a q_s 
\frac{k^4  - 4 k^3  \mu q + k^2(4 \mu^2 q^2  - p^2  - q^2 )}{4p^2 q^2}.
\label{cs}
\end{eqnarray}
The trace of this tensor structure, \ie 
$c_2(k,q) = \sum_a C(k,q,a,a)$, required for the energy spectra 
evaluation, is equal to 
\begin{equation}
c_2(k,q) = \frac{(1-\mu^2) k^2}{4p^2}
\bigl[k^2(d-1)-2kq\mu d + 2q^2(d+2\mu^2-2)\bigr].
\label{c2trace}
\end{equation}
In the important case of $d=3$ this gives  
\begin{equation}
c_2(k,q) = \frac{(1-\mu^2) k^2}{2(k^2+q^2-2kq\mu)}\bigl[k^2 - 3kq\mu +q^2(1+2\mu^2) \bigr].
\label{c2t}
\end{equation}
The integral in frequency argument of the product of squared response 
functions in the integral \eqref{c2eff} gives in the limit of zero frequency  
$k_0\!\to\!0$:  
\begin{equation}
\int_{-\infty}^\infty \frac{dq_0}{2\pi} |G_0(q)|^2 |G_0(k-q)|^2 
\to \frac{1}{2\nu^3} 
    \frac{1}{\vq^2(\vk-\vq)^2} 
    \frac{1}{\vq^2 + (\vk-\vq)^2}.
\end{equation}
 
In the important case of $d=3$, with the Mexican hat wavelet ($n=2$) taken 
for definiteness, the integral over the angle variable $\mu=\cos\theta$ can be 
evaluated analytically for the single band random force \eqref{a0band}. 
The angle integration in \eqref{cefn} gives 
\begin{equation} 
 C_{eff}^{d=3,n=2}(k) = \lambda^2
\frac{k^2 |G_0(k)|^2 4\pi a_0^{8}D_0^2}{2\nu^3} 
\int_0^\infty dq q^4 i_c(k/q), 
\end{equation}
where 
\begin{eqnarray}
i_c(x) &=& \int_{-1}^1 dy e^{-a_0^2 q^2(2+x^2-2xy)}
\frac{(1-y^2)(1+x^2-3xy+2y^2)}{2+x^2-2xy} = F(x)+F(-x) \label{icx}\\
\nonumber F(x)   &=& \frac{1}{8x^5}\Bigl[ 
\frac{e^{-a_0^2 q^2(2+x^2-2x)}}{a_0^8q^8} 
\bigl(-3 - 2a_0^2q^2 (-1-3x+x^2) + 2 a_0^4 q^4 (-1-2x -2x^2+2x^3) \\
     &+&2a_0^6q^6 (2+2x+x^2)\bigr)  
+ 2(4+x^4) {\rm ExpIntEi}\left( -a_0^2 q^2(2+x^2-2x) \right)
\Bigr].
\end{eqnarray}
In the limit of small wave numbers $x=\frac{k}{q}\to0$ this gives  
$$\int_0^\infty dq q^4 \lim_{x\to0} i_c(x) = 
\frac{7}{80} \frac{\sqrt\frac{\pi}{2}}{a^5}.$$
Therefore the effective correlator 
tends to 
$$
C_{eff}^{d=3,n=2}(k\to0) = \lambda^2\frac{7}{40}
\frac{k^2 |G_0(k)|^2 \pi^\frac{3}{2} a_0^3 D_0^2}{\nu^3\sqrt{2}}. 
$$

\end{document}